\documentstyle[12pt]{article}

\begin{document}
\title{\textbf{Dynamo action at recombination epoch of open Friedmann universe spatial sections}}\maketitle
{\sl \textbf{L.C. Garcia de Andrade}-Departamento de F\'{\i}sica Te\'orica-IF-UERJ- RJ, Brasil\\[-0.0mm]
 \paragraph*{Chicone et al [Comm Math Phys (1997)] investigated existence of fast dynamos by analyzing the spectrum kinematic magnetic dynamo. In real non-degenerate branch of the spectrum, the kinematic dynamo operator lies on a compact Riemannian 2D space of constant negative curvature. Here, generalization of Marklund and Clarkson [MNRAS (2005)], general relativistic GR-MHD dynamo equation to include mean-field dynamos is obtained. In the absence of kinetic helicity, adiabatic constant $\gamma=\frac{1}{2}$ and gravitational colapse of negative Riemann curvature of spatial sections enhance dynamo effect $\frac{{\delta}B}{B}=2.6\times 10^{-1}$. Critical time where linear dynamo effects breaks down de to curvature. At recombination time, COBE temperature anisotropies, implies that magnetic field growth rate is ${\lambda}{\approx{10}^{-9}yr^{-1}}$. This places a bound on curvature till the recombination magnetic field is amplified to present value of $B_{0}=10^{-9}G$, by dynamo action. At present epoch, negative curvature becomes constant and the Chicone et al result is shown to be valid in cosmology. Since negative curvature is non-constant, Hilbert theorem which forbiddes negative constant curvature surfaces embeddeding in $\textbf{R}^{3}$ is bypassed.}PACS: 47.65.Md, 02.40-Ky. Key-words: Dynamo plasmas; Riemannian geometry.}
\newpage
\section{Introduction}
The real astronomical universe is certainly turbulent. Either inside black holes or galactic nucleus or even in the early universe, things do not behave in the very simple realm of laboratory physics of low velocity hydrodynamics and cosmology even if relativistic is not so laminar as in the present stages of universe expansion. On the other hand dynamo theory \cite{1} has been one of the most successful theory able to explain galactic \cite{2} and solar magnetism \cite{3}. The mean field dynamo theory developed mainly by Raedler and Krause \cite{4}, paved the way to the better understand of the randomic processes in the turbulent plasma. From the mathematical point of view, the first simple solution of dynamo equation in Riemannian space was given  by Arnold et al \cite{5}, which have made use of a somewhat unrealistic hypothesis of uniform stretching in Riemannian spaces given by steady flows. More realistic Riemannian solutions in pseudo-Anosov space has been presented by Gilbert \cite{6} following the where particles and magnetic field lines were stretched along a torus map. Another sort of fast dynamo as presented by Chicone et al \cite{7} as a two-dimensional compact stretched Riemannian manifold. Chicone et al proved that in order this manifold could host a fast dynamo, the Riemannian curvature had to be constant and negative. Yet more recently Garcia de Andrade \cite{8} has presented two solutions in three-dimensional stretched Riemannian space representing fast dynamos in plasmas, all of these solutions of dynamo Faraday equation are not very realistic because they basically do not take into consideration oscillation in the fields and fluctuations of random fields, which happens naturally in mean field theory as presented by Raedler and Krause \cite{4}. \newline
Investigation of dynamo action on those Riemannian manifolds are fundamental to obtain the connection between dynamo theory and experiment in unusual topological flows as recently shown Moebius strip dynamo flows \cite{9}, in the case of the Perm liquid sodium dynamo experiment. In this last example of slow dynamo, a dynamo wave was obtained along the twist directions of the Moebius dynamo flow. Question why to link Riemannian 2-manifold of constant negative curvature to dynamos, or a Anosov space to dynamos, is responded by the Chicone et al recently work who showed that, fast dynamos in 2D can only be realized in Riemannian spaces of constant negative curvature, without violating Cowling anti-dynamo theorem \cite{10}. In this paper, one obtains expressions for the Faraday self-induction equation in Riemannian space of non constant curvature, starting from a review on plasma dynamos meanfields. The result is applied to Goedel cosmology, to compute the growth rate of magnetic fields in that universe from COBE data. The fact that the negative Gaussian curvature is non-constant, means that Hilbert theorem stating that, negative constant curvature surfaces cannot be embedded in $\textbf{R}^{3}$ is not violated here. Actually Efimov theorem which generalizes Hilbert result, to slow varying curvature cannot be applied here since or curvature variation is time variation and spatial variation as in Efimov non-embedding theorem of negative curvature in higher dimensional manifolds less than five. One of the main difficulties of obtain physical models of fast dynamos in curved manifolds of negative constant curvature is that this 2D cannot be isometrically embedded \cite{10} in $\textbf{R}^{3}$ or even in $\textbf{R}^{4}$, as given in Hilbert Theorem. Actually, Hilbert stated that the negative Gaussian curvature surface can only be embedded in spaces $\textbf{R}^{n}$, for $n\ge{4}$, and even $n=4$ is not even completely clear. Nevertheless, Hilbert theorem does not say anything of non constant negative curvature. Thus one may consider a 2D spatial section of a cosmological manifold embedded in $\textbf{R}^{3}$, which in turn is a 3 manifold spatial spacetime section. This allow us to build a fast dynamo action in the spatial section of Friedmann spatial section of negative curvature. These would be non Anosov spaces \cite{11}.\newline
This paper is organised as follows: Section II addresses the review of meanfield dynamo plasmas the computation of the relation between the curvature of Riemannian space, the magnetic Reynolds number and the universe expansion in the absence of magnetic helicity. In section III future prospects are presented.
\newpage
\section{Fast dynamos at recombination phase in spatial sections of open cosmology}
Arnold \cite{12} first noticed the importance of the sign of the curvature of the manifold of geodesic flows, in order they can diverge and become unstable in the case of negative constant Riemann curvature. Here one addresses the issue of considering the Friedmann metric in the form devised by Carneiro \cite{13}, from a Goedel metric in order that
\begin{equation}
ds^{2}=a^{2}[dt^{2}-(d{\rho}^{2}+sinh^{2}{\rho}(d{\theta}^{2}+sin^{2}{\theta}d{\phi}^{2}))]\label{1}
\end{equation}
is the transformed Goedel like metric in the Friedmann format in spherical coordinates. A 2D spatial section of this metric can be expressed as
\begin{equation}
-dS^{2}=a^{2}[d{\rho}^{2}+sinh^{2}{\rho}d{\theta}^{2}]\label{2}
\end{equation}
where the Gaussian curvature ${\kappa}=-\frac{1}{{a}^{2}}=nonconstant$ and ${a}^{2}[t]=nonconstant$. Though, as pointed out by Kleides et al \cite{14} the isotropy is a strong metric approximation in cosmology in the presence of the magnetic field, since this field introduces some anisotropy. That is the basic reason why Kleides et al used a Bianchi type I type metric to investigate dynamos and gravitational instabilities. The advantage to addopt the Friedmann metric here to investigate dynamo spectra is that it is simpler to solve the eigenvalue problem and Riemannian spaces of negative constant Riemannian curvature is more suitable to consider in isotropic metrics.
Actually Chicone and Latushkin showed that, fast dynamos in 2D can only be supported in 2D if the Riemannian spaces possesses constant negative curvature \cite{7}. On this hyperbolic section one shall now solve the self-induction GR-MHD dynamo equation of Clarkson and Marklund \cite{15}, which can be expressed as
\begin{equation}
\dot{\textbf{B}}(1+\frac{5}{3}{\Theta}{\eta})-(1+\frac{2}{3}{\eta}{\Theta})\frac{2}{3}
{\Theta}\textbf{B}=
{\nabla}{\times}(\textbf{V}{\times}\textbf{B})+{\eta}{\Delta}\textbf{B}-
{\eta}<\textbf{Ric},\textbf{B}>\label{3}
\end{equation}
Here one addopts the force-free dynamo equation
\begin{equation}
curl B={\lambda}_{0}B
\label{4}
\end{equation}
commomly used in galactic dynamos, and the magnetic relation
\begin{equation}
{\Delta}_{flat}B=-curl(curl B)=-{{\lambda}_{0}}^{2}B
\label{5}
\end{equation}
is the force-free Beltrami equation. Substitution of expression (\ref{12}) into expression (\ref{11}) for the dynamo GR-MHD equation, simplifies a great deal of the equation. In the mean field dynamos case there is a relation between the magnetic and electric fields given by
\begin{equation}
\textbf{E}=\textbf{v}\times\textbf{B}
\label{6}
\end{equation}
This  magnetic and velocity fields obey the mean field law
\begin{equation}
curl v{\times}B=\alpha{B}
\label{7}
\end{equation}
Substitution of these expressions
\begin{equation}
\dot{\textbf{B}}(1+\frac{5}{3}{\eta}{\Theta})-[(1+{\eta}\frac{2}{3}{\Theta})\frac{2}{3}
{\Theta}-{\lambda}^{2}{\eta}]\textbf{B}=-{\eta}<\textbf{Ric},\textbf{B}>\label{8}
\end{equation}
Here the Ricci tensor is given on a coordinate-free language by
\begin{equation}
Ric={R_{ij}}dx^{i}{\otimes}dx^{j}
\label{9}
\end{equation}
where $\otimes$ represents here the tensor product. To compte the expression for the Laplacian in Riemannian space below one shall need the Riemann-Christoffel symbols
\begin{equation}
{{\Gamma}^{i}}_{jk}=\frac{1}{2}g^{il}[g_{lj,k}+g_{lk,j}-g_{jk,l}]
\label{10}
\end{equation}
and its trace
\begin{equation}
{\Gamma}_{j}=\frac{1}{\sqrt{g}}{\partial}_{j}[\sqrt{g}]
\label{11}
\end{equation}
here comma denotes the partial derivative with respect to the spatial coordinates. Here $\langle{Ric,\textbf{A}}\rangle$ is given explicitly on a coordinate chart by the inner public
\begin{equation}
\langle{Ric,\textbf{A}}\rangle={R^{i}}_{j}A^{j}
\label{12}
\end{equation}
 By using the Marklund-Clarkson computation of the Laplacian operator ${\Delta}={\nabla}^{2}$, yields \begin{equation}
 {\Delta}B^{i}=-curl(curl)B^{i}=-D^{2}B^{i}+D^{i}(D_{j}B^{j})+2{\epsilon}^{ijk}{\dot{B}}_{j}{\omega}_{k}+
 2R^{ij}B_{j}\label{13}
 \end{equation}
 By considered that all the vectors involved in the mean-field dynamos are parallel (Fermi) transport along the lines of the flow in 3D Riemannian curved spatial section of the cosmological model, thus
 \begin{equation}
 D_{j}B^{k}=0\label{14}
 \end{equation}
 which also implies $D^{2}B^{i}$ vanishes since $D^{2}=D_{i}D^{i}$ and $D^{i}$ is the covariant derivative operator. Thus the expression for the Laplacian of the magnetic field above reduces to
 \begin{equation}
 {\Delta}B^{i}=-curl(curl)B^{i}= 2\frac{1}{\sqrt{g}}{\epsilon}^{ijk}{\dot{B}}_{j}{\omega}_{k}+2R^{ij}B_{j}
 \label{15}
 \end{equation}
 By taking the approximation that the adiabatic cosmic expansion as
 \begin{equation}
 {\Theta}(t)=\frac{1}{[1+2{\gamma}]t}
 \label{16}
 \end{equation}
 where t is the cosmic time, implies that the expansion nonlinear terms can be dropped to simplify matters. Substitution of this expression into the mean-field dynamo equation reads
 \begin{equation}
 \frac{2}{3}{\Theta}+{\lambda}[1+{\eta}{\Theta}]-{\alpha}{\lambda}_{0}=-\frac{2}{3}\dot{\Theta}{\eta}-\kappa(t){\eta}
 \label{17}
 \end{equation}
 where ${\lambda}$ is the magnetic field growth rate and ${\alpha}$ is the kinetic current helicity. Also $\kappa(t)$ is the Riemannian 2D curvature depending on time. Rearranging terms into this last equation yields
 \begin{equation}
 {\lambda}-{\alpha}{\lambda}_{0}+\kappa(t){\eta}+\frac{2}{3}[1+2\eta]{\Theta}+\frac{2}{3}\eta\dot{\Theta}=0\label{18}
 \end{equation}
 This equation can be physically interpreted twofold, first it may be considered, as a ODE for the expansion once curvature is given. First here one shall addopt the second point of view where the expansion is given by the adiabatic relation (\ref{16}) which yields immeadiately the following Gaussian curvature expression
 \begin{equation}
 \kappa(t){\eta}= -{\lambda}+{\alpha}{\lambda}_{0}-\frac{2}{3}[1+2\eta]{\Theta}-\frac{2}{3}\eta\dot{\Theta}
 \label{19}
 \end{equation}
In the absence of kinetic helicity this dynamo equation becomes
\begin{equation}
\kappa(t){\eta}= -[{\lambda}+\frac{2}{3}[1+2\eta]{\Theta}+\frac{2}{3}\eta\dot{\Theta}]\label{20}
\end{equation}
First interesting consequence of this expression is that for ideal plasma cosmology, where resistivity $\eta$ vanishes, the growth rate of the magnetic field yields
\begin{equation}
{\lambda}=-\frac{2}{3}{\Theta}\label{21}
 \end{equation}
In this case chaotic dynamos can be obtained when the universe, undergoes a contracting phase or gravitational colapse $\Theta\le{0}$. Case equal zero is called marginal dynamo case. Actually this case also yields a fast dynamo since by definition
\begin{equation}
lim_{{\eta}\rightarrow{0}}\textbf{Re}{\lambda}({\eta})>0
\label{22}
\end{equation}
means that a fast dynamo can be supported in this phase of Friedmann universe. Other physical interesting impplications of the dynamo cosmic expansion equation is when the adiabatic relation for ${\gamma}=\frac{1}{2}$ which yields
\begin{equation}
\kappa(t){\eta}= -[{\lambda}+\frac{2}{3}[1+2\eta]{\Theta}+\frac{2}{3}\eta\dot{\Theta}]
\label{23}
\end{equation}
The dynamo growth rate is given by
\begin{equation}
{\lambda}= -\kappa(t){\eta}+\frac{4}{3}[1+2\eta][1+2\gamma]^{-1}t^{-1}\label{24}
\end{equation}
Therefore at the present epoch of the universe this equation reads
\begin{equation}
{\lambda}= -\kappa(\infty){\eta}=0\label{25}
\end{equation}
Thus a dynamo action cannot be supported at the present age of the universe. Far away from the Big Bang, for finite times, signs of a relic dynamo exist on the negative Riemann constant curvature, spatial section of Friedmann universe. This confirms the Chicone et al result in the cosmological context. At the recombination epoch where 
\begin{equation}
t_{R}=10^{-13}sec
\label{26}
\end{equation}
the amplification of the magnetic field reads
\begin{equation}
{\lambda}=2.7\times10^{-13}sec
\label{27}
\end{equation}
The self induction magnetic contrast is expressed as
\begin{equation}
\frac{{\delta}B}{B}=2.6\times 10^{-1}\label{28}
\end{equation}
Amplification of the magnetic field till the present vale of $B_{0}\approx{10^{-9}G}$, may be generated from a small amplification of the order of $\delta{B}\approx{10^{-10}G}$. This is well within Zeldovich limit of  $B\ge{10^{-21}G}$. Now let us compute the breakdown time of the linear dynamo action considered here. This can be obtained by considering the expression
\begin{equation}
\frac{{\delta}B}{B}=1=\lambda t_{c}\label{29}
\end{equation}
This yields the following expression for the critical time as $t_{c}\approx{10^{14}}sec$. The critical cosmic expansion of the Friedmann cosmos, where dynamo is considered as marginal is given by ${\Theta}_{c}\approx{10^{13}}sec^{-1}$. In this computation one considered that the cosmic time derivative is approximately zero, which is the same slow expansion considered by Barrow and Tsagas \cite{16} also in open Friedmann universe. 

\section{Conclusions} 
Dynamo action and dark matter are both two distinct features of the universe which shares, a common characteristic which is the constant curvature of the universe spacetime manifold. That is one of the reasons of the importance of being able to investigate the close relation fast dynamos have and may be produce in spatial regions of the curvature universe with negative curvature. The second reason is the demonstration by Chicone et al that fast dynamo action can be obtained in Riemannian compact manifolds of constant negative curvature and geodesic flows. In this brief report one generalizes their idea to open cosmology where the hypothesis of contancy of negative curvature can be relaxed. Previous model of cosmic dynamo, has been proposed by Bassett et al \cite{17} by making use of pre-heating phases of inflationary models. The present model on the other hand introduces some new features as the consideration of marginal dynamos and the presence of fast dynamo in the spatial sections of open Friedmann universe which naturally possesses a negative curvature. The advantage of using Friedmann model as a dynamo lad is that the cosmic flow in the Friedmann cosmology is also naturally geodetic. Enqvist \cite{18} on the other hand also proposed to used small scale random magnetic fields to explain primordial magnetic fields, which favors the isotropic model used here. Enqvist obtained  $B\approx{10^{-20}}G$, here obtains perturbations of seed fields of the order of $10^{-10}G$ . Therefore here one has considered the embedding of variable negative Riemann curvature which may give rise to a constant negative curvature, compact Riemannian manifold section of open cosmological models. 3D Riemannian curved space embedded in pseudo-Riemannian spacetime appears naturally here Einstein-Friedmann cosmology in general relativitistic dynamos context. Recently \cite{19} dynamo plasmas in two-dimensional non-compact Riemannian manifold were obtained, openning new possibilities for further applications in ideal cosmological plasmas.
\section{Acknowledgements}
Several discussions with D Sokoloff, J L Thiffeault, Y Latushkin, A Bradenburg and K Raedler on possibilities of mean field dynamo theory in Riemannian space, are highly appreciated. Special thanks go to Rafael Ruggiero for helpful discussions on embedding theorems in Riemannian manifolds. I also thank financial supports from UERJ and CNPq.
 
  \end{document}